\documentclass[12pt,preprint]{aastex}
\begin{document}

\title {Infrared Emission from the Dusty Disk Orbiting  GD 362, An Externally-Polluted White Dwarf}

\author{M. Jura\footnote{Department of Physics and Astronomy and Center for Astrobiology, University of California, Los Angeles CA 90095-1562; jura, ben, becklin@astro.ucla.edu}, J. Farihi,\footnote{Gemini Observatory, 670 North A'ohoku Place, Hilo HI 96720; jfarihi@gemini.edu}\,\, B. Zuckerman,$^{1}$\,\&\,E. E. Becklin$^{1}$}

\begin{abstract}
We report {\it Spitzer Space Telescope}  photometry between 3.6 ${\mu}$m and 24 ${\mu}$m and spectroscopy between 5 ${\mu}$m and 15 ${\mu}$m of GD 362, a white dwarf with an effective temperature near 10,000 K  that displays a remarkably high concentration of metals in its photosphere and a thermal infrared excess. We approximately reproduce both the infrared continuum and the very strong 10 ${\mu}$m silicate emission feature with a model of an orbiting dusty disk which is  flat out to  50 stellar radii and  warped  between 50 and 70 stellar radii.  The relatively small amount of cold material implied by the weak 24 ${\mu}$m flux argues that the  disk lies within the Roche radius of the star, and we may be witnessing a system where an asteroidal-size body has been tidally destroyed.  If so, determination of the photospheric metal abundances  may measure the bulk composition of an extrasolar minor planet.
\end{abstract}

\section{INTRODUCTION}

Although little is known about the fate of a planetary system when a main-sequence star eventually becomes a white dwarf,  we can now detect  with modern instrumentation indirect signatures
of  disrupted comets and asteroids  in the atmospheres and circumstellar environments
of cool white dwarfs.  Here, we report  infrared observations of GD 362 that strengthen the argument that this star is orbited by  debris from a tidally disrupted asteroidal-size body.

Gravitational settling of  elements heavier than the dominant light gas -- either hydrogen or helium -- 
through the relatively thin outer convective zones of  white dwarfs  occurs  rapidly (Paquette et al. 1986).
Therefore,  the atmospheres of the stars which are cool enough that radiative levitation is unimportant are anticipated to be metal-free.  In agreement with this expectation, ${\sim}$75\% of hydrogen-rich white dwarfs  with effective temperatures
less than 10,000 K have calcium abundances below 10$^{-5}$  solar (Zuckerman et al. 2003).  However, some cool white dwarfs do display atmospheric metals, and these stars are almost certainly externally-polluted.   One of the most remarkable metal-bearing cool white dwarfs is GD 362 which has   
approximately solar abundances of calcium, magnesium and iron (Gianninas et al. 2004). Motivated  by this striking composition,   Becklin et al. (2005) and Kilic et al. (2005) independently
reported ground-based observations showing that GD 362 has an infrared excess  produced by circumstellar dust which is likely to be the source of the external-pollution required to explain the star's photospheric composition.  After G29-38 (Zuckerman \& Becklin 1987), GD 362 is the second single white dwarf discovered to display an infrared excess. 
Here we report
{\it Spitzer Space Telescope} (Werner et al. 2004) observations of GD 362 to  study better its infrared emission and to constrain models for the origin of the circumstellar dust.

A promising model to explain both the infrared excesses and the relatively high metal abundances within both GD 362 and G29-38 is that these stars accrete from  an opaque dust disk, reminiscent of Saturn's rings, produced
by the tidal-disruption of an asteroid (Jura 2003, Becklin et al. 2005).  A somewhat different approach was taken by Reach et al. (2005b) who proposed that the infrared  data for G29-38 can be explained by an optically thin cloud.  These two different models for the infrared excesses around white dwarfs can imply dramatically different  amounts of mass in the circumstellar environment.  If there is an opaque disk, there may be in excess of  ${\sim}$10$^{24}$ g of 
circumstellar dust, comparable to the mass of Ceres, the largest asteroid in the Solar System.     If the disk is optically thin, there may be only ${\sim}$10$^{18}$ g of circumstellar dust, comparable to the mass of a comet,  although there could be a more massive population of large unseen  bodies.  In ${\S2}$ we present the observations while in ${\S3}$ we describe our model.  In ${\S4}$ we discuss our results and in ${\S5}$ we present our conclusions.  

\section{OBSERVATIONS} 
Infrared Array Camera (IRAC; Fazio et al. 2004)
observations of GD 362 were executed 2005 Aug 25
at 3.6 ${\mu}$m, 4.5 ${\mu}$m, 5.7 ${\mu}$m, and 7.9 ${\mu}$m.
A 20 point cycling dither pattern was used with medium scale
steps; 5 frames of 12 seconds taken at each point in the dither
pattern, yielding a total on-source time of 1200 seconds in all
four filters.  At each of the four wavelengths, individual frames were combined into a single
reduced image via the IRAC calibration pipeline, version 12.4.
Aperture photometry was performed on the target in the combined
image using standard IRAF tasks.  Both the flux and SNR were
measured in a 3 pixel (3.6${\arcsec}$) radius, with a sky annulus of 10 - 20
pixels. This measured flux was then extrapolated to the IRAC standard 10 pixel  aperture using 
appropriate corrections (IRAC Data Handbook, Version 3.0, 2006).  For all four channels, the SNR was sufficiently
high ($>$ 25-100) that the photometric errors are dominated by the 10\% absolute
calibration uncertainty of the IRAC instrument that results from the as-yet to be  fully characterized  responses of the filter bandpasses (Quijada et al. 2004, Reach et al.
2005a, Hines et al. 2006).  The results are listed in Table 1; no color corrections have been
applied. 

GD 362 was imaged with the Multiband Imaging Photometer for Spitzer
(MIPS; Rieke et al. 2004) in the 24 ${\mu}$m filter.  The observations consisted of 20 cycles
with the small field of view, the default 14 point dither pattern,
and 10 second individual exposures for a total of 2800 seconds of
on-source integration.  The data were processed through the MIPS
calibration pipeline, version 12.4, to create a single combined
and reduced image.  Standard IRAF aperture photometry was performed
with a 6${\arcsec}$ radius (2.35 pixels) utilizing appropriate sky annuli
and aperture corrections as determined by the MIPS instrument team and instrument support team (http://ssc.spitzer.caltech.edu/mips/apercorr/).  The
measurement yields a SNR = 4.3 and the total error is the
inverse of the SNR summed in quadrature with the 10\% absolute calibration error of the 24 ${\mu}$m camera (MIPS data handbook, Version 3.2, 2006).
The result is listed in Table 1.  From   previous ground-based data (Becklin et al. 2005) and from inspection of the Spitzer IRAC and MIPS images, contamination by other sources of the fluxes we report for GD 362 does not appear to be a problem.  

Spectroscopy over the 5-15 ${\mu}$m region of GD 362 was performed
with the Infrared Spectrograph (IRS; Houck et al. 2004) on 2006
April 18.  The instrument was operated in staring mode at two
positions both along the Short-Low 1 module slit (spectral range between 7.4 ${\mu}$m and 14.5 ${\mu}$m and spectral resolution, ${\lambda}/{\Delta}{\lambda}$, between 64 and 128)   and  along the Short-Low 2 module slit (spectral range between 5.2 ${\mu}$m and 7.7 ${\mu}$m and spectral resolution, ${\lambda}/{\Delta}{\lambda}$, between  80 and 128), with 240 second
exposures executed at each position, repeated twice for a total
on-source time of 960 seconds in each module.  The data were
processed through the IRS calibration pipeline, version 14.0, to
create reduced and combined exposures at each nod position in each
module.  These combined nodded frames were subtracted from one
another to eliminate sky signal and were processed with the data reduction program SPICE  to perform
spectral extraction.  The extracted spectra in each Short-Low module
were averaged to a single spectrum to increase the signal to noise ratio.  The extractions
were performed twice: first with the default aperture function
(8 pixels at 12 ${\mu}$m) in order to assess the proper calibration level,
which does not exist for custom (non-default) aperture extractions;
and second with a smaller aperture function (4 pixels at 12 ${\mu}$m)
which yielded a higher SNR extraction.  The custom extraction was
scaled appropriately to match the overall level of the default
extraction and all the orders were combined (and averaged in
regions of overlap, including the Short-Low 1 bonus section which has a spectral range of 7.3 ${\mu}$m to 8.7 ${\mu}$m).  In our data we  clearly detect only silicate emission; higher signal to noise is required to determine if other features are present. The
spectrum and the associated errors are displayed in Figure 1.

 The {\it Spitzer} data show a strong silicate feature near 10 ${\mu}$m, consistent with the ground-based photometry presented  by Becklin et al. (2005). 
 The profile of the silicate emission from GD 362  resembles the emission feature seen by Reach et al. (2005b) for G29-38, and it  is distinctly different  from  the profile of interstellar silicates (Kemper et al. 2004) also shown in Figure 1.
  If the  dust around GD 362 has an asteroidal origin, its spectrum may resemble that of warm ($T$ $>$ 200 K) dust around main-sequence stars (see Chen et al. 2006).   BD +20 307 notably exhibits  a strong silicate feature from grains at a temperature near 650 K (Song et al. 2005), and  we see from Figure 1 that the spectra of GD 362 and BD +20 307 agree fairly well except that the blue wing of the  feature is stronger in
 GD 362, which, as described below,  can be understood if the grain temperature is near  ${\sim}$1100 K.  
\section{PROPOSED MODEL}

Becklin et al. (2005)  found that their ground-based continuum data from 1.2 ${\mu}$m to 11.3 ${\mu}$m for GD 362 could be moderately well matched with a geometrically-thin, flat, opaque disk whose total thickness is much less than the diameter of the star and where the dust temperature is   a function only of the radial distance from the star.  The flux in the $N'$ (11.3 ${\mu}$m) filter in the MICHELLE instrument on the Gemini North telescope was substantially above the prediction of the model disk.  This discrepancy is now understood as a result of the very strong silicate emission (see Figure 1) entering into the $N'$ filter which has a full-width half-maximum of 2.4 ${\mu}$m.  Without any temperature gradient vertical to the disk, this model cannot, by itself, explain the strong silicate emission.
Therefore, we present a more sophisticated disk model.   

We  assume a passive disk with three regions so 
that at every radius, the disk re-radiates the  flux incident upon it.  A schematic edge-on view of our model is shown in Figure 2.  
 The inner region has sufficient surface density that it is opaque at all wavelengths of interest.  Furthermore,  thermal conductivity is sufficiently great in this region that the disk is vertically isothermal.  Further out, the disk is still opaque, but thermal conductivity is less important, and there is a temperature gradient in the disk. The outer region of the disk
is thick enough to absorb most of  the incident radiation, but it is optically thin in the infrared. This outermost portion of the disk has relatively little mass, and therefore it can
be warped. Our model is somewhat artificial since the three regions are distinct, and therefore it has unrealistic temperature jumps at the boundaries between the regions. These temperature jumps are most important in the midplane of the disk which emits little light and thus the model predictions are not strongly affected by our overly simplified description of the disk temperature variations.   

  If $R$ denotes the distance from the star and
${\tau}_{\nu}$ denotes the optical depth from the top of the disk measured downwards,  then as with a stellar atmosphere, the emergent intensity, $I_{\nu}$, for a  disk seen at inclination angle $i$ defined such that a face-on disk has $i$ = 0$^{\circ}$,  is
\begin{equation}
I_{\nu}(R)\;=\;{\int}_{0}^{\infty}B_{\nu}(T[{\tau}_{\nu}])\,\exp(-{\tau}_{\nu}/\cos\,i)\,\left(d{\tau}_{\nu}/\cos\,i\right)
\end{equation}
If $D_{*}$ denotes the distance to  GD 362, the flux received at  Earth, $F_{\nu}$, is
\begin{equation}
F_{\nu}\;=\;D_{*}^{-2}\,\cos\,i\,{\int}_{R_{in}}^{R_{out}}\,2{\pi}\,R\,I_{\nu}(R)\,dR
\end{equation} 
To model the observed flux, we compute the temperature at all locations within the disk.

\subsection{Region I: The Inner Opaque, Vertically-Isothermal Region}
 Close to the star, unshielded grains can get very hot and rapidly sublimate.  However, if thermal conductivity between a grain and the surrounding gas is important in regulating a grain's temperature, then    the disk can
 be approximately vertically isothermal.  Consequently, even close to the star, grains at the top of
 the disk can survive.  
 
 The importance of thermal conduction depends upon the amount of gas  in the disk. In the Solar System,
 Ceres has a mean density of 2.1 g cm$^{-3}$ and an estimated water content between 17\% and 27\% by mass (McCord \& Sotin 2005, Thomas et al. 2005).   The tidal destruction of an analog to Ceres  could thus provide  the disk with  gaseous hydrogen, and a gas to dust ratio by mass  between 0.02 and 0.03.  Additional  elements such as sodium and potassium can be gaseous at 1300 K in metal-enhanced environments (Ebel \& Grossman 2000), but we ignore this contribution to the gas composition.

 Thermal conduction is important in determining the grain temperature, $T$,  when the rates of energy exchange from the grain surface to its surroundings by gas particles and photons are comparable.  This situation occurs when  (Burke \& Hollenbach 1983):
\begin{equation}
n_{0}\,k_{B}\,T\,\left(\frac{8\,k_{B}\,T}{{\pi}\,m}\right)^{1/2}\;{\approx}\;{\sigma}_{SB}T^{4}\
\end{equation}
where ${\sigma}_{SB}$ is the Stephan-Boltzmann constant, $k_{B}$ is Boltzmanns's constant,
$n_{0}$ is the gas number density in the midplane of the disk and $m$ is the mean atomic weight of the gas.  
 If ${\Sigma}_{gas}$ denotes the mass surface density of the gaseous fraction of the disk and with the usual assumption that the disk is in vertical hydrostatic equilibrium, then at distance $R$ from the star of mass $M_{*}$:
\begin{equation}
n_{0}\;=\;\frac{{\Sigma}_{gas}}{m}\,\left(\frac{GM_{*}m}{2{\pi}R^{3}k_{B}T}\right)^{1/2}
\end{equation}
From equations (3) and (4), thermal conduction is important if:
\begin{equation}
{\Sigma}_{gas}\;{\geq}\;\frac{{\pi}\,m\,{\sigma}_{SB}\,T^{3}}{2\,k_{B}}\left(\frac{R^{3}}{GM_{*}}\right)^{1/2}
\end{equation}
Adopting very approximate values of $R$ ${\sim}$ 10$^{10}$ cm, $T$ ${\sim}$ 1000 K, $M_{*}$ ${\sim}$ 0.8 M$_{\odot}$, and $m$ = 3.3 ${\times}$ 10$^{-24}$ g (for H$_{2}$),  thermal conduction is important where ${\Sigma}_{gas}$ ${\geq}$ 0.2 g cm$^{-2}$.  
 If the disk has a mass of
10$^{24}$ g and a mean radius of  ${\sim}$10$^{10}$ cm (Becklin et al. 2005),  the average dust surface density is
3000 g cm$^{-2}$.  Therefore, in regions where the gas density is $>>$10$^{-4}$ of the grain density as expected from the disruption of an analog to Ceres (see above), thermal conduction is important.  
For a flat disk where the thermal conduction inhibits any vertical thermal gradient, the temperature is (see, for example, Chiang \& Goldreich 1997):
\begin{equation}
T\;=\;\left(\frac{2}{3{\pi}}\right)^{1/4}\,\left(\frac{R_{*}}{R}\right)^{3/4}\,T_{*}
\end{equation}

\subsection{Region II: The Opaque Region with Vertical Temperature Gradients}

Beyond the innermost region, grains at the top of the disk can survive even without efficient thermal conduction, and there may be a vertical temperature gradient.  To compute the grain temperature as a function of location, we adopt a model similar to that  of a planetary atmosphere  Goody \& Yung (1989) or a  disk around a pre-main-sequence star (Calvet et al. 1991).  We define ${\alpha}$ as ratio of the Planck mean opacities in the optical, ${\overline{\chi}_{V}}$ and infrared, ${\overline {\chi}_{IR}}$:
\begin{equation}
{\alpha}\;=\;\frac{{\overline{\chi}_{V}}}{{\overline{\chi}_{IR}}}
\end{equation}
In this approach, we assume that the grains are heated by optical light and re-radiate infrared light.
We assume that all of the dust particles  at a given distance from GD 362 have the same radius, $a$, and therefore at each location a single temperature characterizes the material.  We also assume that the spectral character of the incident stellar radiation does not vary with optical depth so that ${\overline{\chi}_{V}}$ is constant throughout the atmosphere.  We further neglect
the temperature variations in ${\overline{\chi}_{IR}}$.   Consequently, ${\alpha}$ depends only upon the grain size, and it varies as a function of its distance from the star since we allow for different-size grains
at different radial locations.  Below, in \S3.4, we describe in more detail our sources for deriving ${\alpha}$.  
 
If ${\mu}_{0}$ denotes the mean cosine of the average incidence angle of the starlight with respect to the normal to the disk and if we ignore limb darkening, then at distance $R$ from the star:
\begin{equation}
{\mu}_{0}\;=\;\frac{4}{3{\pi}}\,\frac{R_{*}}{R}
\end{equation}
where $R_{*}$  denotes the radius of the star.  
Integrated over all frequencies, the mean intensity of the incident optical light, $J_{opt}(0)$, at the top of the disk is:
\begin{equation}
J_{opt}(0)\;=\;\frac{R_{*}^{2}}{8\,R^{2}}\,\frac{{\sigma}_{SB}T^{4}_{*}}{{\pi}}
\end{equation}
 For convenience, we define $J'(0)$ as:
\begin{equation}
J'(0)\;=\;\frac{1}{8}\,\left(\frac{R_{*}}{R}\right)^{2}
\end{equation}

To find the grain temperature as a function of depth, we compute the mean intensity of the optical light within the
atmosphere, J$_{opt}({\tau})$, from:
\begin{equation}
J_{opt}({\tau})\;=\;J'(0)\,\left(\frac{{\sigma}_{SB}T_{*}^{4}}{{\pi}}\right)\exp{\left(-\frac{{\alpha}\,{\tau}}{{\mu}_{0}}\right)}
\end{equation}
Assuming that the infrared source function is $({\sigma}_{SB}\,T^{4}/{\pi})$,  an approximate solution to the equation of transfer for the temperature as a function of depth is (see Goody \& Yung 1989):
\begin{equation}
\left(\frac{T}{T_{*}}\right)^{4}\;=\;J'(0)\left(\left[{\alpha}\,-\,\frac{3\,{\mu}_{0}^{2}}{\alpha}\right]\,exp\left(-\frac{{\alpha}{\tau}}{{\mu}_{0}}\right)\;+\;\left[2\,{\mu}_{0}\,+\,\frac{3\,{\mu}_{0}^{2}}{{\alpha}}\right]\right)
\end{equation}
Since we expect that  ${\alpha}$$>$ 1 and ${\mu}_{0}$ $<<$ 1, then at the ``top" of the atmosphere where ${\tau}$ = 0:
\begin{equation}
T(0)\;{\approx}\;\left(\frac{{\alpha}}{8}\right)^{1/4}\,\left(\frac{R_{*}}{R}\right)^{1/2}\,T_{*}
\end{equation}
This result for $T(0)$ is that of a black body  that  is fully illuminated by the half of the star which is visible above the opaque disk.  Deep in the disk where ${\tau}$ $>>$ 1, we find from equations (8), (10) and (12) that
the temperature asymptotically approaches the value:
\begin{equation}
T({\infty})\;{\approx}\;\left(\frac{1}{3{\pi}}\right)^{1/4}\,\left(\frac{R_{*}}{R}\right)^{3/4}\,T_{*}
\end{equation}
This asymptotic temperature is 0.84 of the temperature predicted in the vertically isothermal disk, as given by equation (6).  

\subsection{Region III: The Outer Optically Thin Region}
At a large enough distance from the star, the disk surface density is sufficiently small that the system is optically thin in the infrared.    To compute the temperature in this region, we note that the disk may also be
warped, as described in the Appendix.  
For simplicity,  we define ${\mu}_{warp}$ as the mean warping angle, and analogous to equation (11), the mean intensity
of optical  light in the disk is:
\begin{equation}
J_{opt}\;=\;J'(0)\,\left(\frac{{\sigma}_{SB}T_*{}^{4}}{{\pi}}\right)\exp\left(-\frac{{\alpha}{\tau}}{{\mu}_{warp}}\right)
\end{equation}
To compute the grain temperature, we use:
\begin{equation}
\frac{{\sigma}_{SB}\,T^{4}}{{\pi}}\;=\;{\alpha}\,J_{opt}
 \end{equation}
Assuming that ${\mu}_{warp}$ $<<$ 1, we   use a modified version of equation (1) to estimate the intensity from the warped portion by integrating to the optical depth where  $J$ falls to 50\% of its value at the top of the warped disk instead of to infinite optical depth.   The azimuthal variation of the warp is
not constrained by the data; only the mean warping angle.

\subsection{Results}

We now describe a specific model of the system to fit the data.   Gianninas et al. (2004) derived a high gravity and mass for GD 362 with the assumption that
the atmosphere is largely composed of hydrogen.  A more recent study of the spectrum suggests that the Balmer line profiles are better reproduced if there a significant amount of helium in the atmosphere of the star (Garcia-Berro et al. 2006).  Independently, Zuckerman et al. (2006) have detected an absorption line at 5876 {\AA} which is strong, direct evidence for the presence of helium.  Consequently, the stellar parameters derived by  Gianninas et al. (2004) need to be revised in the sense that their values for the inferred  gravity and mass of the star  are too large and their values for the inferred radius and distance to the star are too small.  However, for the purposes of modeling the infrared emission from the star, we need only the effective temperature and the ratio of $R_{*}/D_{*}$.  Here, we adopt $R_{*}/D_{*}$ = 5.2 ${\times}$ 10$^{-12}$, consistent with the distance and radius assumed by Becklin et al. (2005).  In the infrared, the spectral shape of the star's photosphere is taken as a single-temperature black body and is independent of the distance to the star.   Based on IRAC observations of other white dwarfs (Kilic et al. 2006), this approximation is usually accurate to better than the 10\% measurement uncertainty in the fluxes for white dwarfs with effective temperatures warmer than 7000 K (Tremblay \& Bergeron 2006).  Therefore, we follow Becklin et al. (2005) to separate the photospheric and
circumstellar contributions to the infrared light.
At 3.6 ${\mu}$m,  the photosphere contributes about 30\% of the total flux, and much less at longer wavelengths.   In using equation (2), we assume a face-on disk.

Given that we have detected a silicate emission feature,  we can constrain the size and composition of the dust particles.  
Reach et al. (2005b) fit the observations of G29-38 with a combination of amorphous silicate (to account for the peak of the feature) and
forsterite (to explain its red wing).   Here, for simplicity, to derive ${\alpha}$, we  include only glassy silicate grains  with an olivine stoichiometry  described by Dorschner et al. (1995) for the case with equal abundances of magnesium and iron.  As shown below, we find a satisfactory fit to the silicate feature
without any forsterite because we include relatively large grains.  Our model for the grains is similar to that used by Chen et al. (2006) to account for the silicate emission from the debris disk which emits near 360 K around the main sequence star ${\eta}$ Crv.

We adopt model parameters to match the data; a summary of the dimensions and temperatures of our model disk is given in Table  2.      Following Gianninas et al. (2004), we assume that the effective temperature of the star is 9740 K, but this value may be revised upwards perhaps as much as ${\sim}$10\% with
a better measure of the atmospheric helium abundance.  We assume that the temperature of the grains is  less than approximately 1200 K so that they do not rapidly sublimate; this choice for the maximum grain temperature  is consistent with the spectral energy distribution of the infrared excess excess near 3.6 ${\mu}$m.  Given this maximum grain temperature, we adopt an inner disk radius of 12 $R_{*}$.  
The relatively low flux at 24 ${\mu}$m means that the disk
cannot extend too far outwards, for otherwise there would be more cool dust and therefore more
emission at this wavelength than measured.   We therefore assume that the opaque disk is truncated at
50 $R_{*}$.  To fit the entire infrared continuum,  we  adopt a transition between  Region I and  Region
II  at 30 $R_{*}$, although the overall fit is insensitive to this choice and Region I could extend between 12 $R_{*}$ and 50 $R_{*}$.    In order that the grains at the top of the disk in Region II not exceed 1200 K, we adopt ${\alpha}$ = 1.5 which is achieved by our model glassy grains with radii of 2 ${\mu}$m.

In our model, the silicate emission is largely produced in Region III, the optically thin outer portion of the disk beyond 50 $R_{*}$.  So that the grain temperature lies below about 1200 K, we adopt  ${\alpha}$ = 5 which is
characteristic of  our model glassy grains with a radius of 1 ${\mu}$m. 
To enable the very strong silicate feature to reprocess about 1\% of the total luminosity  of the star, we assume Region III is warped because a flat disk cannot intercept enough light from the illuminating star. The maximum fraction of the star's luminosity that a flat disk
of inner radius $R$ can re-emit is $(4R_{*})/(3{\pi}R)$, and with $R$ = 50 $R_{*}$,  the outer portion of the disk can reprocess at most 1\% of the energy of the star.  Since the dust might
emit only ${\sim}$50\% of its light in the silicate feature,  the outer portion of a flat disk cannot
account for the strength of the feature.    
There is a degeneracy between the warp angle and the outer radius of region III of the disk.  Since the grains that produce the silicate emission cannot also produce much emission at 24 ${\mu}m$, we infer
that these particles are relatively warm. We therefore, adopt 70 $R_{*}$ as the outer radius  of the warped thin disk.  With this outer radius, the strength of the silicate feature is reproduced with
a mean warp angle of 7.5$^{\circ}$.

 We display in Figure 3 a comparison 
between the data and the reasonably-fitting  model.   
  We somewhat overestimate the  flux at 24 ${\mu}$m, but this measurement is only a 4${\sigma}$ result and therefore somewhat uncertain.  Also, the width of the peak in the model spectrum is somewhat narrower than that shown by the data so that other materials besides glassy silicates might be present in the disk.    Crystalline material may contribute to the shape of the silicate emission, and  a spectrum longward of  15 ${\mu}$m can test this possibility (Jaeger et al. 1998).  Regardless of these uncertainties, it seems that the dust
  lies fully within the star's tidal radius of ${\sim}$1 $R_{\odot}$, consistent with the hypothesis that the dust disk around GD 362 was created by tidal disruption of an asteroidal-size parent body.

\section{DISCUSSION}

Our results constrain possible models for the origin of the dust near GD 362.   
For optically thin emission, the mass of  circumstellar dust, $M_{dust}$, is:
\begin{equation}
M_{dust}\;=\;\frac{F_{\nu}\,D_{*}^{2}}{{\chi}_{\nu}\,B_{\nu}(T)}
\end{equation}
where F$_{\nu}$ is the flux.  Consider  the silicate feature. The peak flux is about 1.5 mJy and with $T$ = 1000 K (see Table 2) and  ${\chi}_{\nu}$ = 3000 cm$^{2}$ g$^{-1}$ (Dorschner et al. 1995), then $M_{dust}$ ${\sim}$  3 ${\times}$ 10$^{17}$  g if the distance is 25 pc as assumed by Becklin et al. (2005). Since it is likely that the star is further away than previously assumed, the mass in dust is likely to be at least 10$^{18}$ g.    If, however,  the atmosphere has a convective envelope characteristic of helium-rich white dwarf at 10,000 K of ${\sim}$1 M$_{\oplus}$, (MacDonald, Hernanz \& Jose 1998) and if 0.1\% of this mass is composed of metals (Gianninas et al. 2004), then the star  may have already accreted more than 10$^{24}$ g.   Consequently,  it is plausible that there is much more circumstellar mass  than inferred from the silicate emission feature.  Supported by  our modeling of the infrared spectral energy distribution, we suggest that the bulk of the circumstellar
mass resides in an opaque, flat disk.

Garcia-Berro et al. (2006) proposed that GD 362 is the result of the merger of two lower mass white dwarfs and that the circumstellar disk is a remnant of this event (see also Livio et al. 2005).  One motivation for this model was
that, according to Gianninas et al. (2004), the mass of GD 362 is 1.24 M$_{\odot}$.  However, with the detection of helium in the photosphere,  the gravity and mass of this star are smaller
than previously derived and it seems much  less likely that  GD 362 is the product of a merger.   In any case, even young massive white dwarfs do not characteristically display infrared excesses from dust (Hansen et al. 2006).   A test of the merger model for GD 362  is to measure the composition of the
star's atmosphere (Zuckerman et al. 2006) to compare with the nucleosynthetetic predictions for such an event.    

The relatively weak 24 ${\mu}$m flux is best understood if there is  little
cold dust in the system.  Interstellar accretion  results in matter drifting radially inwards from the Bondi-Hoyle accretion radius which is typically greater than 1 AU (see Spitzer 1978).  In our model,  the dust is all located
within 0.01 AU.  A similar argument applies to the dust orbiting G29-38 for which Reach et al. (2005) measured F$_{\nu}$(24 ${\mu}$m) = 2.4 mJy.  If the opaque, geometrically-flat disk described by Jura (2003) extends to much more than approximately 1 R$_{\odot}$, then the predicted flux is near 16 mJy.  We therefore interpret the data for G29-38 as evidence that the disk around this star has an outer truncation within 1 R$_{\odot}$.  It is possible that there is cold dust  much further from the star that
is unrecognizable  with current data.  

The model that the dust is located in an orbiting disk is dynamically plausible and is consistent with the infrared observations.  However, we do not have any direct measurements of the disk dynamics or
geometry.  An indirect measurement that constrains the dust geometry in G29-38, a white dwarf that bears many similarities to GD 362, is given by the periodic intensity variations in the infrared disk emission which are driven
by the non-radial  optical-light variations of this ZZ Ceti class star (Graham et al. 1990, Patterson et al. 1991).  These timing data exclude the possibility, raised by the optically thin model of Reach et al. (2005), that  G29-38's circumstellar dust  is distributed in a spherical shell.  If GD 362 is variable (see Gianninas, Bergeron \& Fontaine 2006 for a recent discussion of ZZ class stars), then the spatial distribution of the circumstellar dust could be probed by measuring time-variations of the infrared excess.  

It is possible that the mass in dust orbiting GD 362 substantially exceeds 10$^{20}$ g and therefore the diameter of the tidally-distorted
 parent body was considerably larger than 30 km.  With available data, we cannot determine whether this parent body was more analogous to an asteroid or to a Kuiper Belt Object.  Jura (2006) has argued that  values of $n(C)/n(Fe)$ in three externally-polluted helium-rich white dwarfs  are at least a factor of 10 below Solar,  strongly favoring accretion of asteroid-like bodies with a chondritic 
composition.  Measurements of the   carbon abundances in the atmospheres of G29-38 and
GD 362   may strongly
constrain the nature of the  parent body pollution in these and other metal-enriched white dwarf atmospheres (Zuckerman et al. 2003).

\section{CONCLUSIONS}
  
  We report {\it Spitzer Space Telescope} observations of GD 362.   We approximately  reproduce both the infrared continuum and the strong silicate emission  with a  model  consistent with the  hypothesis that the dusty disk around GD 362 is the debris from   a tidally disrupted asteroidal-size body.  The weak 24 ${\mu}$m flux argues against much cold dust being present in the system as would be expected if
  interstellar accretion is important.  
  Determination of the metal abundance within the atmosphere of this star (Zuckerman et al. 2006) may
  indirectly measure the bulk composition of an extrasolar minor planet.

This work has been partly supported by NASA.

 \appendix
 \begin{center}
 {\bf APPENDIX}
 \end{center}
 \renewcommand{\theequation}{A\arabic{equation}}
 \setcounter{equation}{0}
Following Pringle (1991), we present a simple  description of a dust disk  to assess whether it can be warped.    We assume that the disk is formed from an  asteroidal-size body that is initially disrupted at its periastron, $R_{per}$ and relatively quickly forms a thin ring  which then expands  under the action of its own viscosity to form an orbiting disk with mass $M_{disk}$.  After the initial conditions damp out, the surface density, ${\Sigma}$, is conveniently expressed (Jura et al. 2002) as:
 \begin{equation}
{\Sigma}\;=\;\frac{M_{disk}}{(12{\pi}^{3}\,R^{3}\,v_{vis}t)^{1/2}}\,\exp\left(-\frac{R}{3v_{vis}t}\right)
\end{equation}
where $v_{vis}$ is the radial flow speed due to viscosity and $t$ is the age of the disk.   We define ${\epsilon}$ as:
\begin{equation}
{\epsilon}\;=\;\frac{v_{vis}\,t}{R_{per}}
\end{equation}
so that  ${\epsilon}$   measures the ratio of the spreading distance of the disk compared to
its initial location.  Because the value of $v_{vis}$ is not well measured, we use ${\epsilon}$ to characterize a mature disk.   If ${\epsilon}$ $<<$ 1, then the disk has not had much time to evolve from its initial configuration while if ${\epsilon}$ $>>$ 1, the disk may have largely dissipated.   Therefore,  we adopt ${\epsilon }$ ${\sim}$ 1 for the systems of interest.  

  While a warp might be driven by hydromagnetic (Quillen 2001) or magnetic (Lai 1999) torques, we consider radiative-driven warping.  According to Pringle (1996), if the disk is opaque in
the infrared, it is unstable to self-induced warping  if the star's luminosity, $L_{*}$,  is large enough so  the backreaction of radiation leaving the surface of the disk can induce a sufficiently strong torque, which occurs if:
\begin{equation}
L_{*}\;>\;12{\pi}^{2}{\nu}_{2}\left(\frac{GM_{*}}{R}\right)^{1/2}\,{\Sigma}
\end{equation}
where ${\nu}_{2}$ is the azimuthal viscosity.  Making the uncertain extrapolation that the azimuthal and radial viscosities are equal, we write ${\nu}_{2}$ = $R\,v_{vis}$, and then using equation (A1),  warping occurs where:
\begin{equation}
R\;>\;\frac{(12{\pi}{\epsilon}R_{per}GM_{*})^{1/2}}{L_{*}\,t}\,c\,M_{disk}\,\exp\left(-\frac{R}{3{\epsilon}R_{per}}\right)
\end{equation}
  For GD 362  the relevant parameters are very uncertain.  For illustrative purposes, we adopt $L_{*}$ = 2 ${\approx}$ 10$^{-3}$ L$_{\odot}$, $M_{*}$ = 0.8 M$_{\odot}$,  and $M_{disk}$ = 10$^{24}$ g.   Given that the accretion rate might be ${\sim}$10$^{11}$ g s$^{-1}$ (Becklin et al. 2005), we estimate $t$ = 10$^{5}$ yr, ${\epsilon}$ = 1 and
  $R_{per}$ = 3 ${\times}$ 10$^{10}$ cm.  In this case,  the  disk is  warped for $R$ larger than  ${\sim}$50 R$_{*}$.
  
  The disk is unstable to radiative warping only where it is optically thick. 
  We assume that such warping occurs, and then the disk then expands further
  and becomes optically thin.  In this very outermost region, the decay time for the warp can be as long
  as the disk expansion time (Armitage \& Pringle 1997), and therefore the optically thin portion of the disk effectively  retains the warp it acquired when it was more compact.  

\clearpage

\begin{table}
\caption{{\it Spitzer} Photometry of GD 362}
\begin{tabular}{cl}
\hline
\hline
Wavelength &  Flux \\
(${\mu}$m )& (mJy) \\
\hline
3.6 & 0.380 ${\pm}$ 0.038\\
4.5 & 0.395 ${\pm}$ 0.040\\
5.7 & 0.425 ${\pm}$ 0.043\\
7.9 & 0.644 ${\pm}$ 0.064\\
24 & 0.22 ${\pm}$ 0.06 \\
\hline
\end{tabular}
\end{table}
\clearpage
\begin{table}
\caption{Model Parameters}
\begin{tabular}{lllllllll}
\hline
\hline
Region & $R_{min}$ & $T_{top}$ & $T_{mid}$ & $R_{max}$ & $T_{top}$ & $T_{mid}$ & ${\alpha}$ \\
  & ($R_{*}$) & (K) & (K) & ($R_{*}$) & (K) & (K) \\
\hline
I & 12 & 1025 & 1025  & 30 & 520 & 520 & \\
II & 30 & 1170 & 430 & 50 & 910 & 260 & 1.5 \\
III & 50 & 1220 & 1030 & 70 & 1040 & 870 & 5 \\
\hline
\end{tabular}
\end{table}

\clearpage

\begin{figure}
\plotone{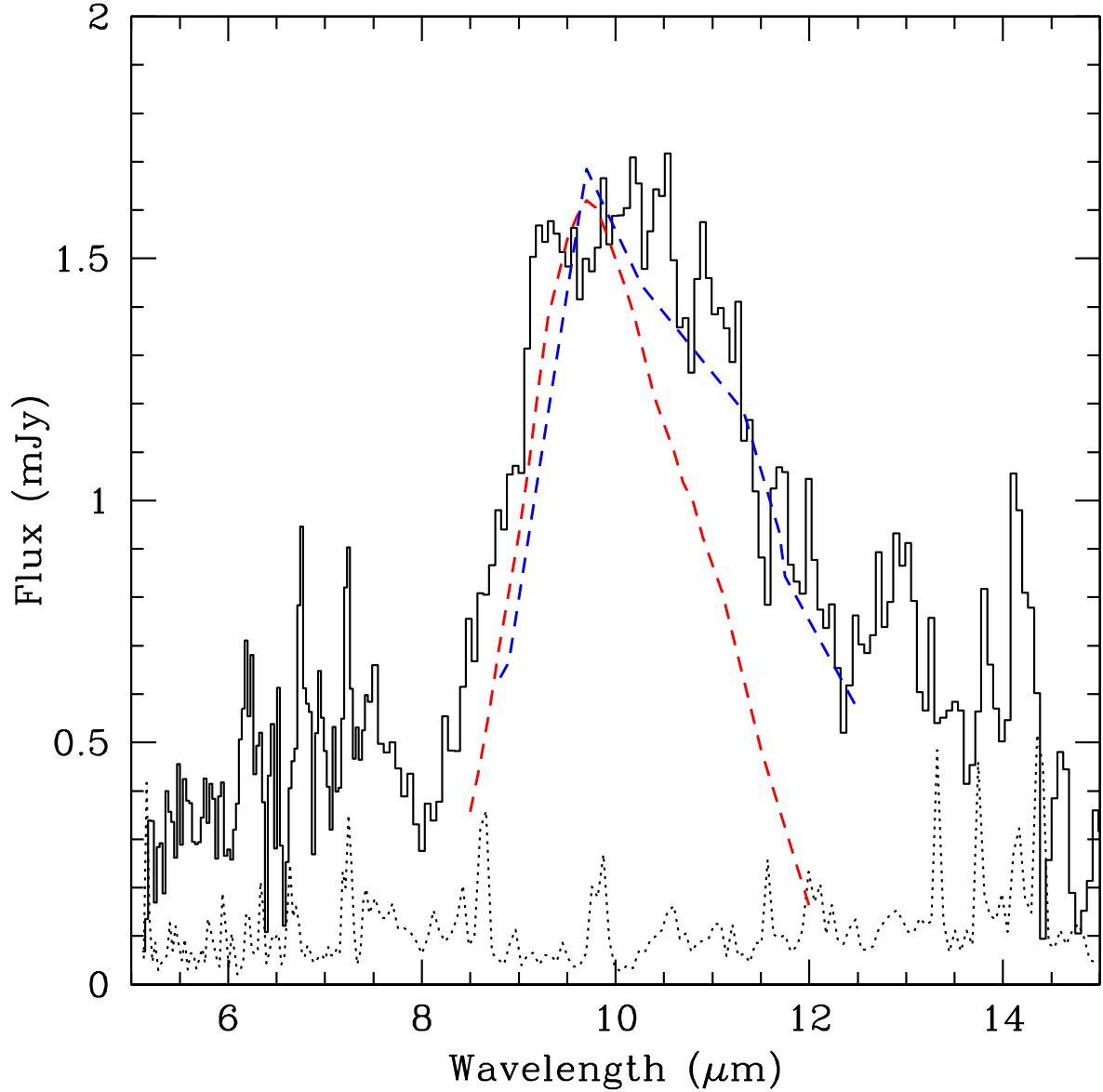}
\caption{IRS spectrum of GD 362.  The errors are shown as the dotted black line; the feature near 14 ${\mu}$m is a detector/instrument artifact.  For comparison, we display the 
 scaled profiles of interstellar silicates (Kemper et al. 2004) as a dashed-red line and the emission from BD +20 307 (Song et al. 2005) as a dashed-blue line.}
\end{figure}
\begin{figure}
\plotone{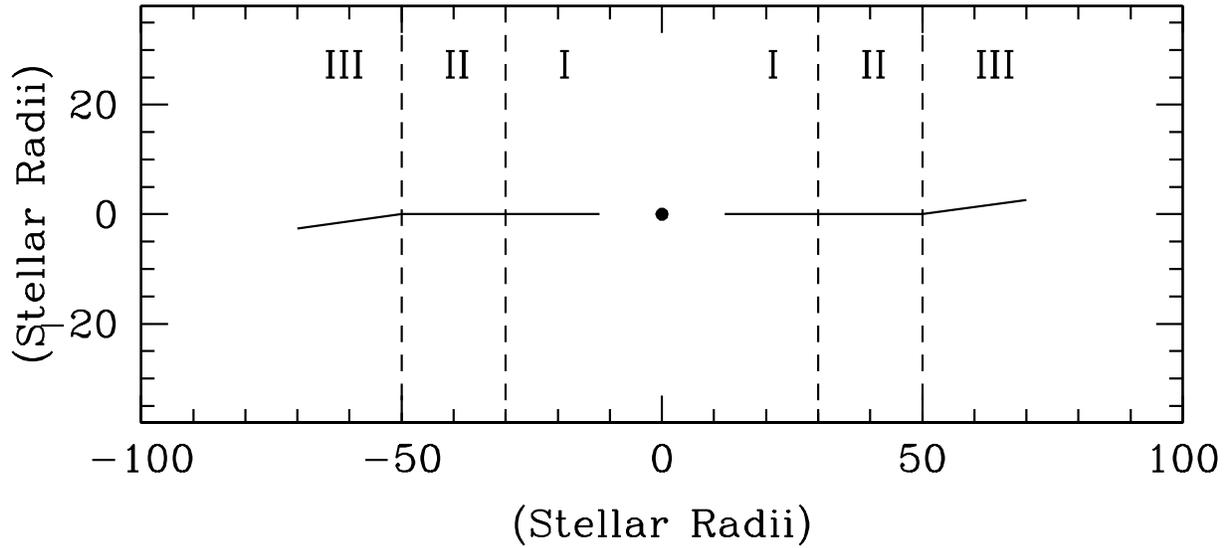}
\caption{Schematic diagram of an edge-on view of our model for the disk orbiting GD 362.  The star is denoted by the small point (approximately to scale) at the origin while the different zones of the disk are labelled.  The azimuthal variation of the warping is not known; here we just illustrate one possibility.}
\end{figure}
\begin{figure}
\plotone{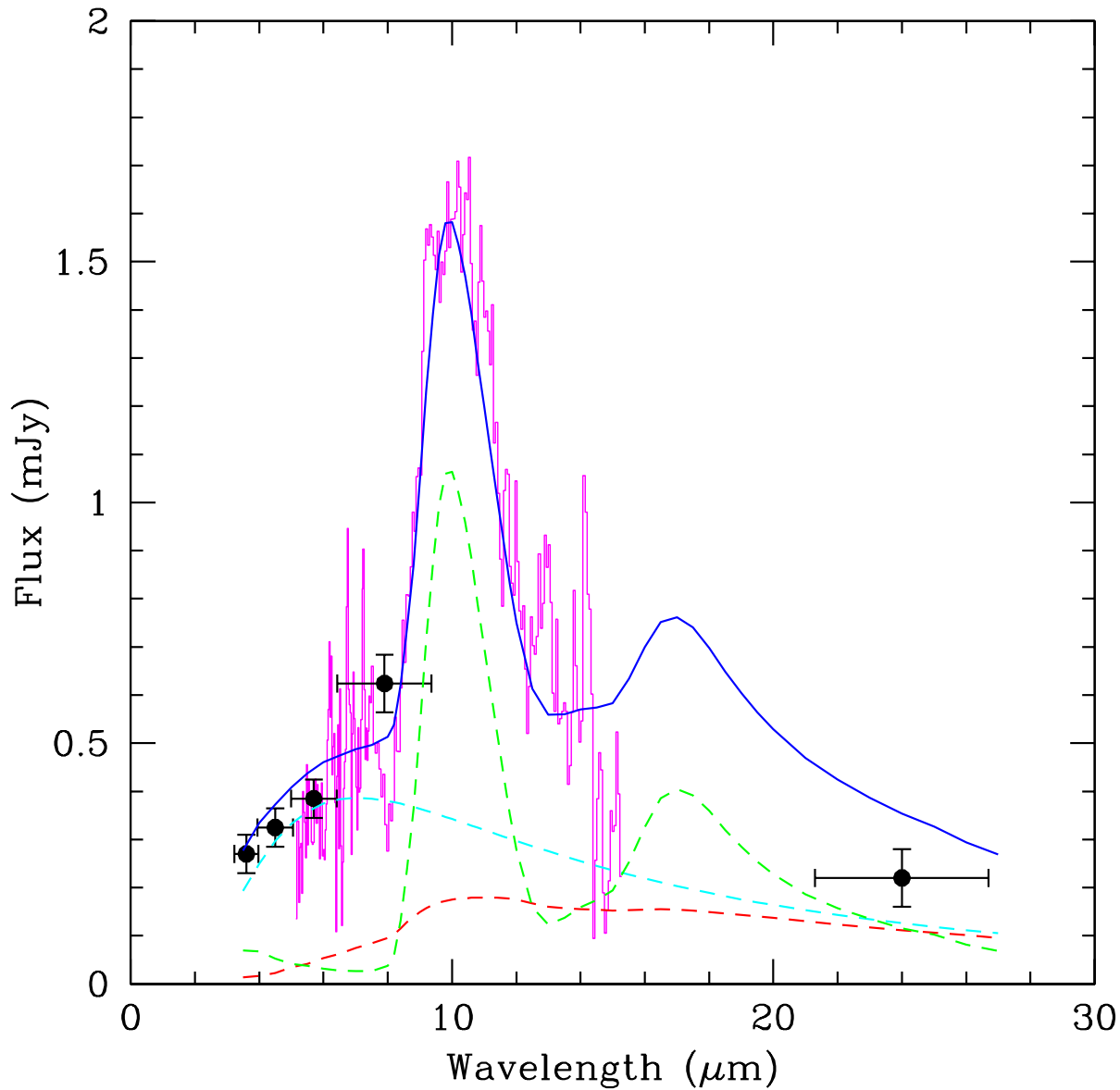}
\caption{Comparison between photospheric-subtracted observations and predicted fluxes for the model disk described in the text.  The solid points represent the IRAC and MIPS data listed in Table 1 while the IRS data from Figure 1 are shown as the magenta line.  The solid blue curve shows the total flux from the model while the fluxes from Regions I, II and III are shown as  dashed lines of cyan, red and green, respectively. }
\end{figure}

\end{document}